# A review of hydrodynamic investigations into arrays of ocean wave energy converters

S. De Chowdhury[1], J.-R. Nader[2], A. Madrigal Sanchez[1], A. Fleming[2], B. Winship[2], S. Illesinghe[1], A. Toffoli[1], A. Babanin[1], I. Penesis[2] & R. Manasseh[1,3]

## Abstract

Theoretical, numerical and experimental studies on arrays of ocean wave energy converter are reviewed. The importance of extracting wave power via an array as opposed to individual wave-power machines has long been established. There is ongoing interest in implementing key technologies at commercial scale owing to the recent acceleration in demand for renewable energy. To date, several reviews have been published on the science and technology of harnessing ocean-wave power. However, there have been few reviews of the extensive literature on ocean wave-power arrays. Research into the hydrodynamic modelling of ocean wave-power arrays is analysed. Where ever possible, comparisons are drawn with physical scaled experiments. Some critical knowledge gaps have been found. Specific emphasis has been paid on understanding how the modelling and scaled experiments are likely to be complementary to each other.

## Contents

---

[1] Centre for Ocean Engineering, Science and Technology; Swinburne University of Technology; VIC 3122; Australia.
[2] Australian Maritime College/ National Centre for Maritime Engineering and Hydrodynamics; Launceston Tasmania 7250; Australia.
[3] Corresponding author (email: rmanasseh@swin.edu.au, Phone: +61392148929).

# 1 Introduction

## *1.1 The promise of wave power*

Wind constantly blows over the 70% of the Earth's surface that is ocean, transferring energy to thesea surface which is averaged out over thousands of kilometres, and delivered to us in the form of ocean swell.

The present availability of wave power worldwide is shown in Figure 1. The International Energy Agency predicts that if nations adopt an emission reduction target to stabilise greenhouse gas emissions in the atmosphere to 450 parts per million of $CO_2$ equivalent, marine energy technologies are likely to grow 14.6 % annually until 2030 (Geoscience Australia, 2010).

In 1974, Salter described ocean wave energy as a clean, safe, permanent and relatively simple source of energy that is yet to be harnessed to its full potential (Salter, 1974) This statement is equally true today, after more than four decades of development.Wave energy has advantages over other forms of Ocean Renewable Energy (ORE), such as tidal energy, ocean thermal energy conversion and ocean current power, since it has a high energy density and low dependency on the local weather. Ocean waves could supply 2000 TWh/year (Chen et al, 2013), sufficient to cover 10 percent of the electricity used globally in 2005. The other forms of ORE have diverse constraints such as insufficient power production apart from a few locations (tidal), having low cost-effectiveness (ocean thermal energy conversion) and having low energy density (ocean currents) (Tseng, et al 2000). Moreover, tidal energy projects have faced strong opposition owing to environmental-impact concerns (CSIRO, 2012). A major review of Wave Energy Converter (WEC) technology was made by Falcão, (2010), and the details of WEC engineering will not be discussed in the present paper.

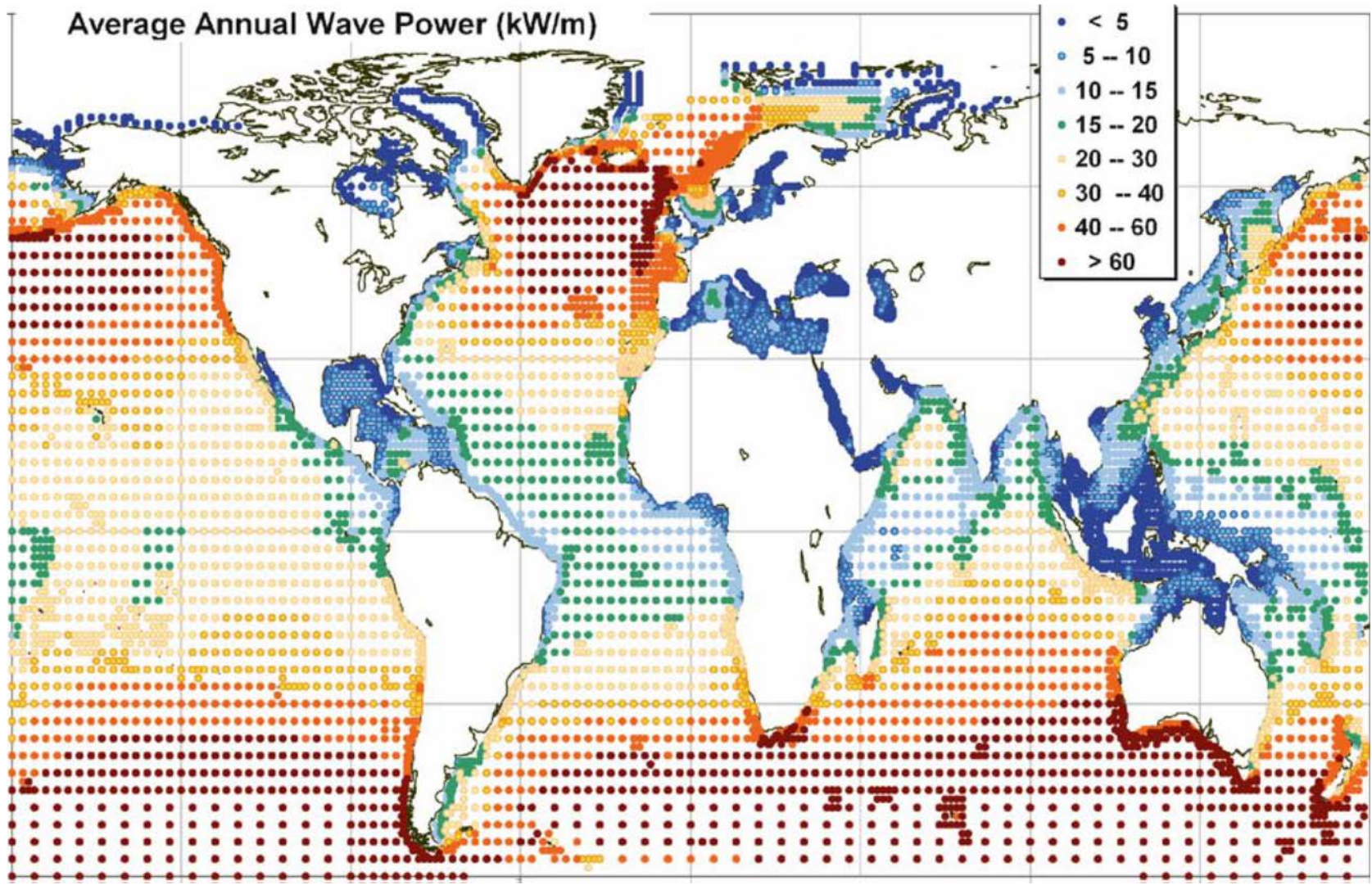

*Figure 1. Global annual mean wave power estimates in kW/m (data from the ECMWF WAM model archive; calibrated and corrected by Fugro OCEANOR against a global buoy and Topex satellite altimeter database), taken from Cruz, 2007.*

The rest of this introductory section 1 discusses issues of WEC operation in arrays in general terms appropriate to a non-specialist in ocean wave or ocean wave power; section 2 briefly discusses the history of individual WECs; section 3 describes the history and current state of the art of ocean deployments of WEC arrays; section 4 reviews the general principles behind the mathematical theories used since the 1970s to model WEC operation in arrays, both analytically, and numerically; section 5 details the numerical simulations of arrays; section 6 reviews physical simulations of arrays; and section 7 provides a brief summary.

## *1.2 Advantages and Issues of wave energy converters operation in arrays*

There are several advantages and potential issues related to wave energy converters operation in arrays. In the following, economic advantages, environmental uncertainties and the array design problem is briefly discussed although the latter will be further examined in section 4.

### 1.2.1 Economic advantages

There are several economic advantages to have WECs set up in 'farms'. Firstly, array interactions, if properly designed,can result in higher power production compare to the same isolated number of WECs; these will be further discussed in the present review. In addition,WEC operation in arrayscouldreduce cost of production when multiple units are built instead of one. There is also a clearcapital cost reduction in power-transmission infrastructure by developing grid connection hub. Moreover, maintenance of multiple WECs will be facilitated due to their relative proximity. And finally, installing different types of WECs or designing them for different sea states couldminimise power-grid fluctuations(Flocard&Finnigan, 2010).

### 1.2.2 The environmental uncertainties:

Any structure at sea, especially in relatively shallow water, will impact the local hydrodynamics of the site in terms of currents, wave amplitude and therefore sediment transport. Some potential issues are listed below.

- Visual or aesthetic impacts.

- Localised silting and un-silting effects,limited due to the radiated property of the wave and decreasing with irregular wave.

- Potential shoreline protection due energy absorption.

- Impact on the localecology, which may be negative or positive since marine structuresmay offersanctuary for marine life as have . tourism benefits etc.

The interaction of wave energy and the shoreline creates local currents, such as rip current (Dalrymple & Lozano, 1978) by the process of transfer of wave energy into local current (Babanin, 2011). In general, currents bring nutrients and organisms from the deep to sustain nearshore marine life (Reommich& McGowan, 1995) and also remove pollutants (Rodriguez et al, 1995), sustaining the whole ecosystem and local fishing industries (Fulton & Bellwood, 2005). Such local currents created by waves are also responsible for transporting sediments, eroding some zones along the shore and building up sandbars and beaches elsewhere (Roy et al, 1994). Furthermore, wave action can re-suspend sediment, which influences local marine biology (Larcombe et al, 2001). Significant alteration to the wave regime by wave-power arrays could have a detrimental environmental and knock-on regional economic impact.

In simple terms, the local currents created by waves are proportional to the power of the waves (proportional to the square of the wave height) and also proportional to the gradient of the wave height with distance (Riley, 2001). If something causes incoming waves to lose height abruptly with distance (such as the waves passing through an effective array of wave-power machines), there would be strong local currents generated in the immediate vicinity of the array.

The coupling and scattering physics detailed in (1.2 a)) above is *linear* physics, but the environmental-current physics is inherently *nonlinear*. This means that we cannot simply measure the environmental-current physics in a model and scale numbers up to compare with the full scale.

### 1.2.3 The array design problem

Unlike wind or tidal turbines, where the flow is mostly impacted in the wake of a device, WECs affectthe wave field all around them and therefore impacts the environment of all other nearby devices (Simon,1982; Mei, 2012). There are two main issues in array designs, the positioning issue and the coupling issue which can be directly related to thediffraction property of the waves and the radiation properties of WECsand diffraction property of the waves.

**The radiation problem:**

The radiation coupling effect is symbolized in Figure 2.Ocean waves come from left to right. (a) Waves impact the first machine (yellow dot); (b) the first machine is then set into motion in order to absorb energy. This motion creates a radiated wave; (c) both the incoming waves and the radiated waves impact the second machine, so its response is therefore dependant of the motions of the first machine; (d) the second machine starts radiating its own waves, impacting the first machine and altering its response to the incoming waves.It follows that, variation of one machine radiation wave will impact all the other devices in the array. Each machine is therefore coupled to each other. The radiation properties of a device is dependent on its Power Take Off (PTO) characteristics which can be changed in order to harness maximum energy depending on the array configuration and wave properties. But this coupling demonstrate that PTO power optimisation of an array needs to take into account all the devices and be applied to the overall array system. Optimisation applied to one device only may detriment all the other devices and the overall power absorbed. Methods to optimise all the PTO characteristics in array exist and will be further described in chapter 3.

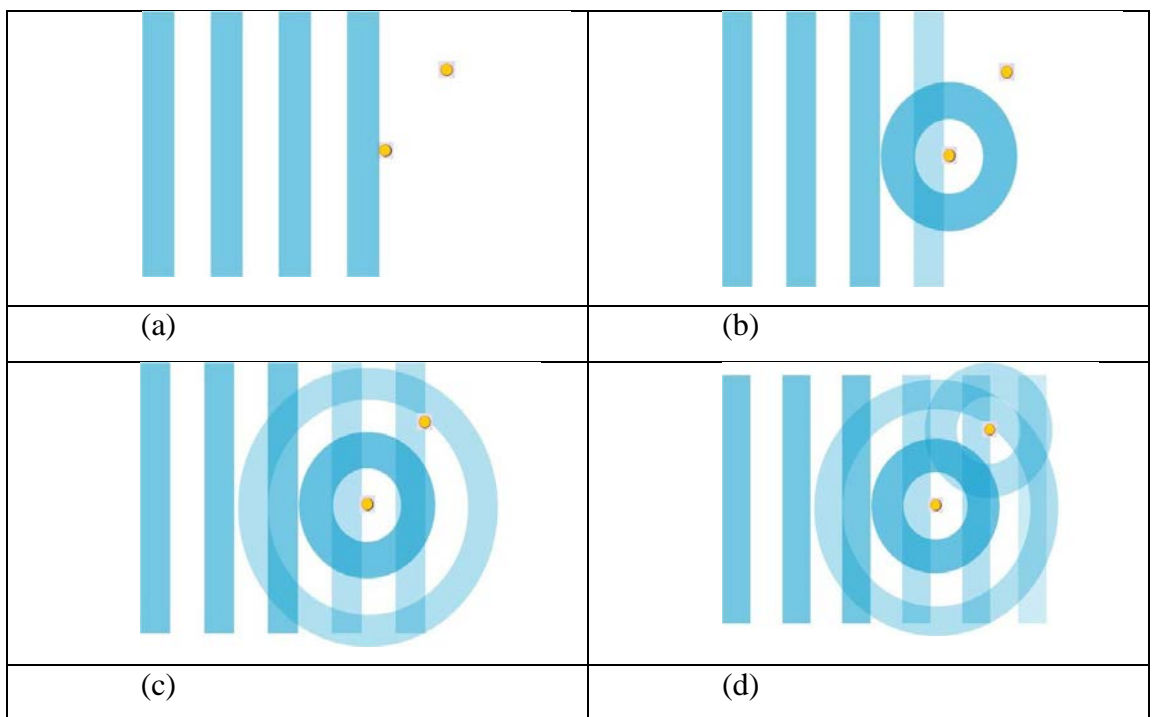

*Figure 2 Radiation problem and coupling between wave machines. Ocean waves come from left to right. (a) Waves impact the first machine (yellow dot); (b) the first machine is then set into motion in order to absorb energy. This motion creates a radiated wave; (c) both the incoming waves and the radiated waves impact the second machine, so its response is therefore dependant of the motions of the first machine; (d) the second machine starts radiating its own waves, impacting the first machine and altering its response to the incoming waves.*

**The diffraction problem:**

Thepositioning problem derive from the diffraction properties of the waves. Figure 3 present the amplitude of the wave field due to diffraction of the incident wave around a single cylindrical OWC device and three type of array arrangements as from (Nader et al., 2014)).

The diffraction wave create the excitation force or volume flux providing the core available energy for the PTO system to absorb. As illustrated in Figure 3, the diffraction field differs strongly depending on the arrangement and similarly will the related excitation components differ. Certain arrangement can induce an average higher excitation force than for a single device, allowing higher available energy known as constructive interaction. Whereas other arrangement can induced lower excitation force than for a single device, destructive interaction. Such effect give rise to the arrangement or positioning optimisation issue. This is a more difficult problem than for the coupling system but methods will also be discussed in Section 3.

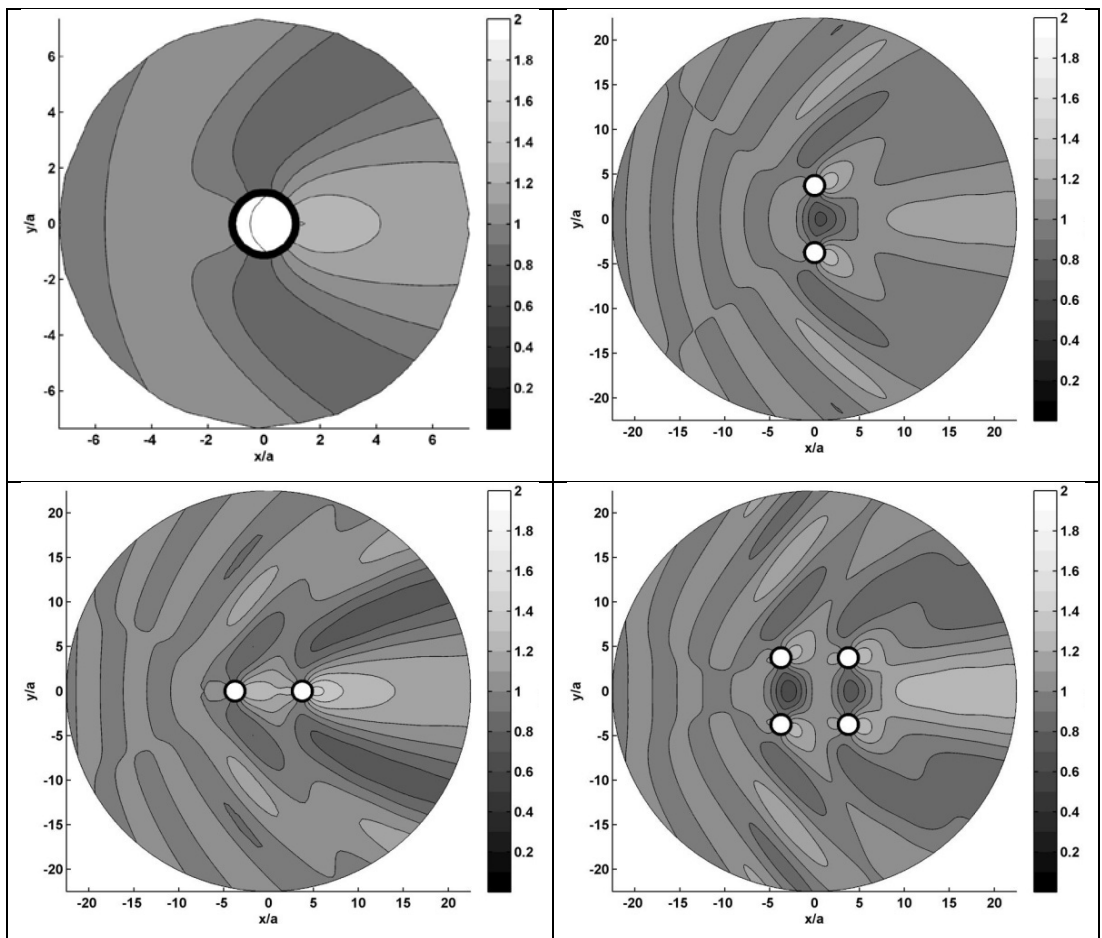

*Figure 3 (From (Nader et al., 2014)) Diffraction around different arrangement of cylindrical OWC devices.*

## 2  Present Scenario in Wave Energy Extraction in Arrays

Owing to the practical issues of operations in the ocean, the deployment of WEC systems requires a significant initial cost. Such an endeavour is usually supported by a close collaboration between public (government) and private sectors. The present scenario in WEC extraction worldwide is thus manifested by different policies adopted by governments in each respective country. (Abdmouleh et al., 2015).

Wave Energy Converters include various systems for capturing energy such as overtopping, oscillating bodies and oscillating water columns (for a review on the existing WEC technologies see, Falcão, 2010). It is important to mention that high pressure can be generated within some WECs through the use of pump type Power Take Off (PTO) systems. These devices can directly produce fresh water from the ocean by reverse osmosis (CSIRO, 2012). WECs could therefore be an indirect groundwater supplier (Sharmila et al, 2004).

The most commonly used classification of WECs considers their effect on the incident waves, thus classing WECs as *point absorbers*, *attenuators*, and *terminators* (Falçao, 2010) One can also group the existing prototypes into two types based on their fundamental mode of operation: i)radiation type and ii) dissipative type. Most WECs as heaving buoys, pendulum or flap-type devices, oscillating water column and some attenuator type devicesabsorbs energy through motion or surface pressure creating a radiated wave as discusses previously. On the other hand, an*overtopping* device or an array device located oncoastrely onlocal wave dissipation due to breaking.

Interest in the development of WEC arrays is growing. To date, however, there have been no full-scale field studies designed specifically to determine how individual machines interact, depending on array layout, distance between machines, seafloor features, incident wave energy regime including direction and variability, and tides and currents including rips and longshore transport.

A brief description of WEC array operations follows.

### 2.1  *Pelamis(Wave Energy Scotland)*

Pelamis Wave Power (PWP) was founded in 1988. The first full scale Pelamis device was installed in 2004 at the European Marine Energy Centre in Orkney, Scotland, UK. The first array of Pelamisdevices were installed in 2008 in Aguçadoura wave farm in Portugal. This was the first ever operating WEC array. Details of this project and subsidiaries have been reported in Dalton et al (2010). These initial successes led to the development of a second generation device, P2. Presently the P2 remains in Orkney and future development is under exploration.In January 2015 the PWP intellectual property and other assets were acquired by the company Wave Energy Scotland (WES). This company was established in December 2014 as part of the Highlands and Islands Enterprise, and is fully funded by the Scottish Government.

## 2.2 Aquamarine Wave Power

This company's technologyemerged in 2001 from the research and development team at Queen's University, Belfast, UK, where researchers worked on flap-type wave energy devices. In subsequent years, these lead to the development of the Oyster technology (Figure 4). With the support of industrial entities Scottish Southern Energy and Sigma capital group, in 2009 they installed the first full scale device named, Oyster 1, at the European Marine Energy Centre (EMEC) in Orkney[4].

In 2011, Aquamarine installed the next generation Oyster 800 device at EMEC. The Oyster 800 is particularly suitable for wave energy farms (Sarkar et al., 2014). Among its notable successes are the survival of the units during the strongest storms of the decade. An array consisting of 40-50 Oyster WEC devices has been proposed for the Isle of Lewis, Scotland, UK (Aquamarine Power Limited 2012).

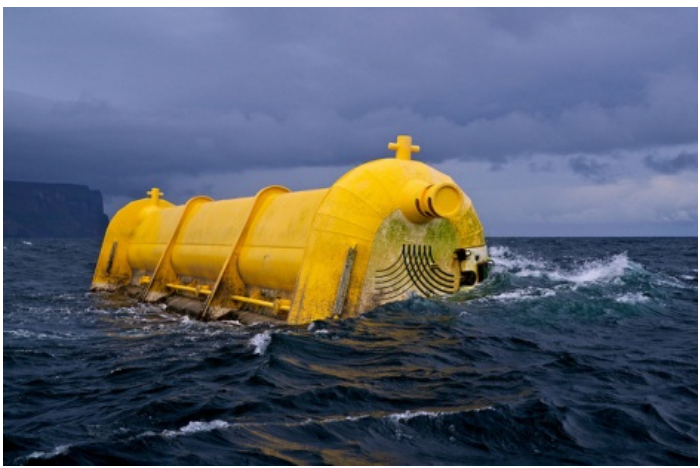

*Figure 4 Aquamarine Power's Oyster 800 wave energy machine.*

## 2.3 Carnegie Wave Energy

This company based in Western Australia has developed a technology named CETO (Figure 5). In contrast to the above-noted technologies, this WEC device is kept below the free surface. The dynamic pressure imparted by progressive waves on the device pumps fluid to shore to run hydraulic turbines.

---

[4]http://www.aquamarinepower.com

Research on CETO began in 1999. The first prototype technology named CETO 1 was developed in 2006, followed by the development of CETO 2,in 2008 and then CETO 3. The CETO 3 device has a power rating of 80 kW. From 2011 onwards, Carnegie embarked on the installation of an array of three devices at the Garden Island site in WA. At the time of writing, two of the three units were still operating, with one having been successfully recovered for scheduled maintenance. This deployment aims at providing both power and desalinated water. Connection to the power gridtook place in February 2015 on Garden Island. The company's next generation device, CETO 6,is designed to integrate power generation within the CETO buoy itself, and is believed to have a power output of 3 MW.

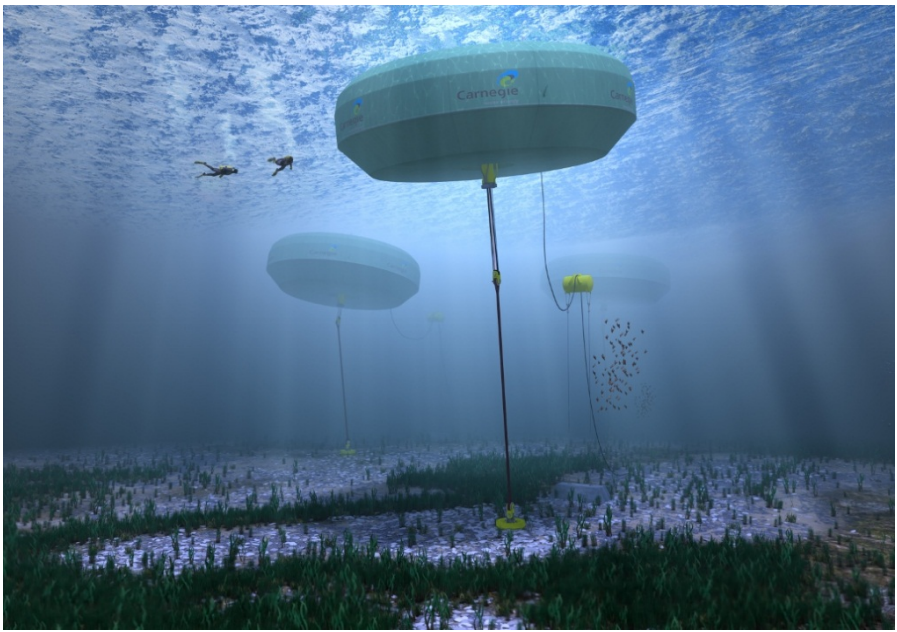

*Figure 5The CETO wave energy device in an array*

The CETO technology has also been tested by other groups like Energies Nouvelles (EDF EN) and French DCNS[5].

## 2.4 Oceanlinx

Oceanlinx was founded in 1997 as Energetech Australia Pty Limited. Major projects took place off Port Kembla, New South Wales, Australlia and were planned for Port McDonnel, South Australia. Oceanlinx held a variety of patents for wave energy extraction. These are Oscillating Water Colume (OWC) devices which harness wave energy by rotation of a turbine due to the movement of compressed air above the water column inside the device. One such device, named blueWAVE, is shown in Figure 6. Under favourable conditions, it had a rated power output 3 MW[6]. In 2010, the maritime certification authority Det Norske Veritas (DNV) validated its power output to be 2.5 MW.

---

[5] http://www.carnegiewave.com

Following a failure in towing its latest device to its intended locationon 2 March 2014, finances forced Oceanlinx to cease operations[7].

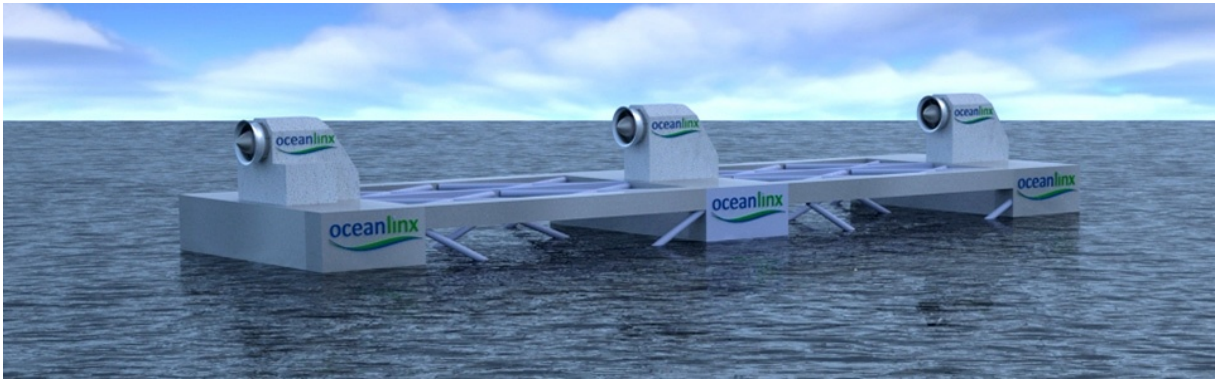

*Figure 6TheblueWAVE device developed by Oceanlinx*

Similar WEC array activities are summarized in Table 1. The above-mentioned companies Pelamis Wave Power, Aquamarine Power and Carnegie Wave Energydescribed above are still retained in Table 1 for comparison. It can be seen that regardless of the WEC technologies employed, mostdevelopers are now planning wave farms based on their existing prototypes of single devices.

*Table 1 Summary of present status of array of WEC devices of few companies*

| Company | Origin | Technology Product | Operation mode | Power Output | Successful installation in array |
|---|---|---|---|---|---|
| AWS Ocean Energy Ltd. | UK | AWS- III | Multi cell array, point type membrane absorber | 2.5 MW | 1:9 scale model in 2010 in Loch ness. |

---

[6] http://www.oceanlinx.com

[7] Climate Spectator, 2014 (2 April), http://www.businessspectator.com.au/news/2014/4/2/renewable-energy/oceanlinx-goes-bankrupt-owing-10m, accessed 11 July 2015.

| Biopower systems Ltd. | Australia | bioWAVE O- Drive | Bottom hinged flap type | 250 kW, 1MW while in wave farm | Under development. |
| --- | --- | --- | --- | --- | --- |
| Embley Energy Ltd. | UK | SPERBOY | Oscillating water column | 450 kW mean power for a single device. | A farm consisting 10 devices is under development. |
| Wave Star APS | Denmark | WAVESTAR | Point absorbers connected to a one way bearing shaft through movable arms | Mean upto 72.71 kW with a $q$ factor 1.14 for an inline array of 5 units. | Under development (Experiments on scaled models completed). |
| OWEL offshore wave energy Ltd. | UK | OWEL WEC | Oscillating water column | 2 MW for a triple duct commercial platform | The multi-duct platform is expected by 2016. |
| ALBATERN | UK | WaveNET | Multi-point absorber | 10 kW for prototype device, 22 kW for array | Isle of Muck, Scotland. WaveNET series (6, 12, 24, series 24 has output 0f 100 MW) for array are being proposed. |
| Ocean | USA | PowerBuoy | Surface | Upto 350 | Under |

| Power Technology | | APB 350 PB 40 | buoy | W for APB 350 and 15 kW for PB 40. | development. |
| --- | --- | --- | --- | --- | --- |
| Carnegie | Australia | CETO | Submerged buoy | 3 MW for CETO 6 device. | Garden Island, WA. |
| Aquamarine | UK | Oyster 800 | Flap type | 40 MW in array | Isle of Lewis in Scotland |
| Pelamis Wave Power | UK | Pelamis | Attentuator | 750 kW for a single device | Aguçadoura, Portugal. |

# 3 Hydrodynamics of WEC arrays

Most studies on WECs and WEC arrays are based on linear water wave theory in the frequency domain. The general assumptions of this theory are inviscid, irrotational and incompressible fluid and the assumption of linearity means that the validity of the theory is limited for small wave amplitudes compare to the water depth and wave length. This theory has been largely and successfully applied for structures at sea in the past. However, WEC research in one sense has the opposite aim to previous marine-structure research. High amplitude motions and resonance need to be avoided at all costs for marine structures; but these are sought after for WECs in order to harness the maximum wave energy. This clearly goes against the limitation of small amplitude in the linear theory and non-linear effects are expected to become significant in WEC arrays. The use of linear theory is, however, a natural first step in solving the coupling and positioning problems and this theory can give some first qualitative results. It is already a quite complex and difficult problem to solve as we will see in the following.

The overall hydrodynamic and phenomenological theory was developed by (Falnes and McIver, 1985), combining the hydrodynamic theory of floating structures (Falnes, 1980) and the pressure-distribution theory by (Evans, 1982). The theory and applications were further developed, reviewed and explained in (Falnes, 2002). This theory is certainly the core of WEC array interactions and will be presented and extended in the following.

## 3.1 The boundary Value Problem

We consider a system of $N_i$ floating bodies and $N_k$ containing an air chamber with an oscillating pressure about a mean value. The system is considered contained within an imaginary cylindrical control surface $S_\infty$. Wetted surfaces of oscillating bodies are indicated by $S_i$ and $S_{i'}$, whereas $S_k$ and $S_{k'}$, denote internal water surfaces. Fixed surfaces, including the sea bed are given as $S_b$, and $S_0$ denotes the external free water surface. The arrows indicate unit normal $n$.

A Cartesian coordinate system $(x, y, z)$ with its corresponding cylindrical coordinates $(r, \theta, z)$ is situated with the origin at the mean sea water level and the $z$-direction pointing vertically upwards. A monochromatic plane wave of amplitude $A$ and frequency $\omega$ propagates from $x = -\infty$ (See Figure 7). Linear water-wave theory is assumed and with the assumptions of irrotational and inviscid flow, a velocity potential $\Phi(x,y,z,t)$ exists that satisfies the Laplace equation

$$\nabla^2 \phi = 0. \tag{Eq1}$$

Under these assumptions $\Phi$ can be expressed using its corresponding complex value $\varphi$ as

$$\phi = \text{Re}\{\varphi(x,y,z)e^{-i\omega t}\} \tag{Eq2}$$

resulting into

$$\nabla^2 \varphi = 0. \tag{Eq3}$$

Inside the fluid domain, a set of homogeneous and inhomogeneous boundary conditions hold.

Homogeneous:

$$\left.\frac{\partial \varphi}{\partial n}\right|_{z=0} = 0 \tag{Eq4}$$

on $S_b$.

$$\left[-\omega^2 \varphi + g \frac{\partial \varphi}{\partial z}\right]_{S_0} = 0 \tag{Eq5}$$

on $S_0$.

Inhomogeneous:

$$\left.\frac{\partial \varphi}{\partial n}\right|_{S_i, i=1,2,..n} = u_{i.n} \tag{Eq6}$$

on $S_i$.

$\boldsymbol{n}$ is the unit vector originating from the surface of the device and pointing outwards.

$$\frac{\partial \varphi}{\partial z} - \frac{\omega^2}{g}\varphi = -\frac{i\omega}{\rho}p_k \tag{Eq7}$$

on $S_k$.

$u_{i,n}$ is the complex normal component of the velocity of body $i$ and $p_k$ is the complex oscillating pressure inside the chamber of body $k$.

The velocity potential can be decomposed into two terms,

$$\varphi = \varphi_D + \varphi_R. \qquad \text{Eq8}$$

where $\phi_d$ is the diffracted wave potential effect of the interaction between the incident wave and the array when they are considered fixed and with constant atmospheric pressure in the chambers. $\phi_r$ is the radiated potential and can further decomposed into

$$\varphi_r = \sum_{ij} \varphi_{ij} + \sum_{k} \varphi_k \qquad \text{Eq9}$$

where $\varphi_{ij}$ is the radiated wave induced by the $j^{th}$ motion velocity $u_{ij}$ ($j = 1, 2, \ldots, 6$ relates to surge, sway, heave, roll, pitch and yaw, respectively) of the body $i$ when no incident wave, other motions or oscillating pressure are present in the system. $\phi_k$ is the radiated induced by the oscillating pressure $p_k$ inside the chamber of the body $k$ when no incident wave, other oscillating pressure or motions are present in the system.

## 3.2 The Phenomenological Theory

The waves create a hydrodynamic force, $F_{i'j'}$, acting on the $j'^{th}$ motion of the body $i'$ as well as a volume flux $Q_{k'}$ induced by the oscillation of the free surface inside the chamber of the body $k'$. $F_{i'j'}$ and $Q_{k'}$ can be discretised following the effect of different velocity potential,

$$F_{i'j'} = F_{i'j',d} + \sum_{ij} F_{i'j',ij} + \sum_{k} F_{i'j',k} \qquad \text{Eq10}$$

and

$$Q_{k'} = Q_{k',d} + \sum_{ij} Q_{k',ij} + \sum_{k} Q_{k',k} \qquad \text{Eq11}$$

In linear theory, there are proportionality between the amplitude of the source of wave, $A$, $u_{ij}$ and $p_k$ and its effect, $F_d$ and $Q_d$, $F_{ij}$ and $Q_{ij}$, $F_k$ and $Q_k$ respectively. Eq. () and () can be rewritten

$$F_{i'j'} = f_{i'j',d} A - \sum_{ij} Z_{i'j',ij} u_{ij} - \sum_{k} H_{i'j',k} p_k \qquad \text{Eq12}$$

And

$$Q_{k'} = q_{k',d} A - \sum_{ij} H_{k',ij} u_{ij} - \sum_{k} Y_{k',k} p_k \qquad Eq13$$

$f_{i'j'}$, $Z_{i'j',ij}$, $H_{i'j',k}$, $q_{k',d}$, $H_{k',ij}$ and $Y_{k',k}$ are complex hydrodynamic coefficients, namely the excitation force coefficient, the radiation impedance, the motion-pressure coupling coefficients, the excitation volume flux coefficient, the radiation admittances and the pressure-motion coupling coefficients. It is noticeable that the real and imaginary parts of $Z_{ij,ij}$ can be directly related to the well-known radiation damping and added mass.

These coefficients, for a given array system, are only dependant on the frequency and direction of the incident wave. Moreover, a set of reciprocity relationship exit,

$$\begin{cases} Z_{i'j',ij} = Z{ij,i'j'} \\ Y_{k',k} = Y_{k,k'} \\ H_{ij,k} = -H_{k,ij} \end{cases} \qquad Eq14$$

These coefficients are the fundament of most of the hydrodynamic and energetic studies of WEC arrays. The reciprocity relationship is therefore an important property as only half these interacting coefficients needs be obtained in order to study the system.

## 3.3 Power absorption

The time-averaged power absorbed by the system from the waves is

$$P_{tot} = \frac{1}{2} \text{Re} \left\{ \sum_{ij} F_{ij}, u_{ij}^* + \sum_{k} p_k Q_k^* \right\} \qquad Eq15$$

Most of the paper included linear properties of the PTO characteristic included in the equation of motion and volume flux pressure relationship. By use of these linearized relationship and Eq. Eq12 and Eq13, $P_{tot}$ can be expressed in function of the hydrodynamic coefficient, PTO damping coefficient and wave amplitude only. It follows that deriving the hydrodynamic coefficients, $P_{tot}$ can be obtain for any desire wave amplitude and PTO damping. This expression also allows to apply direct optimisation methods to obtain the PTO damping for maximum power absorbtion $P_{tot}$.

# 4 The Different Methods for Solving the Boundary Value Problems

## *4.1 Analytical solutions*

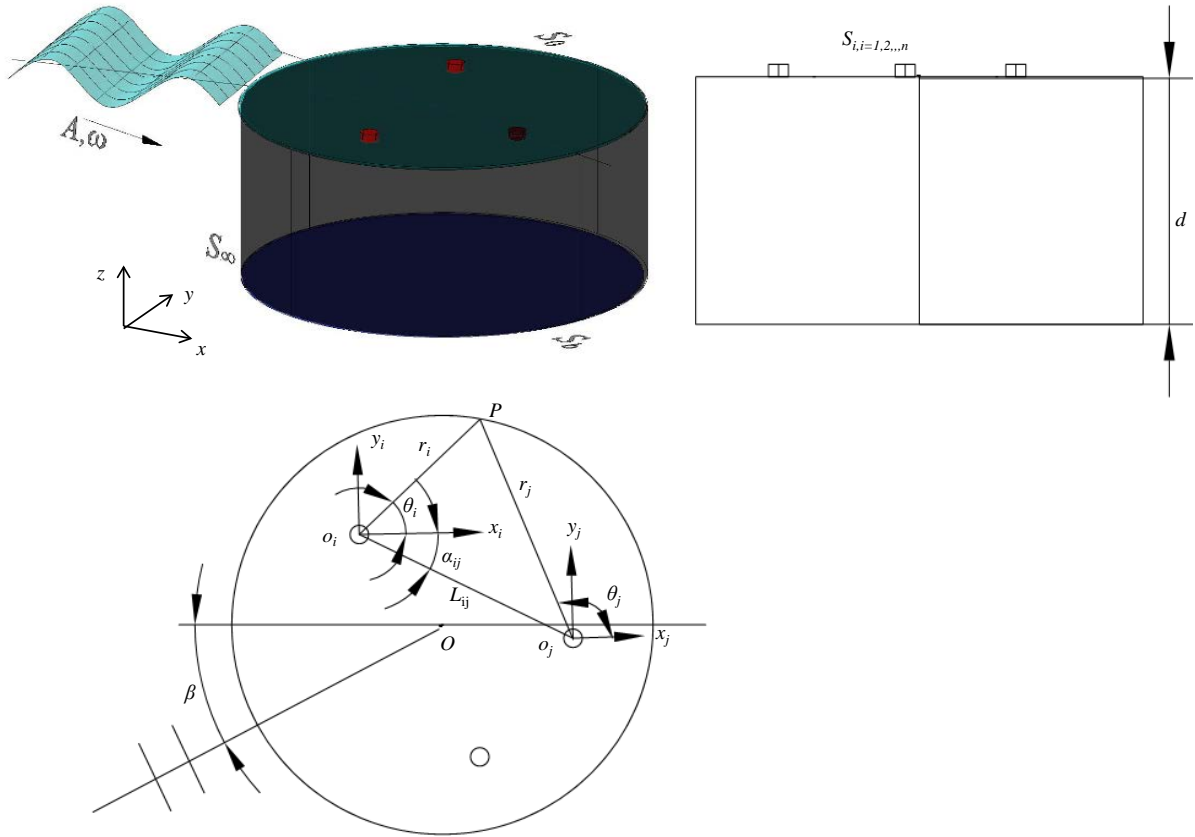

*Figure 7 Schematic of the problem domain used for the analytical modelling of the WEC array.*

In general, as quoted in Mavrakos and McIver (1997), there are three different theories that model WEC arrays analytically: a) Multiple Scattering (MS) approach; b) the Plane Wave (PW) theory; and c) the Point Absorber (PA) theory.

### 4.1.1 Multiple Scattering (MS)

Twersky (1952) proposed a general scheme for solving wave potentials for the case of mutually exchanging system of scatterers. Ohkusu (1974) applied this technique to solve for the wave diffraction due to floating cylinders. Subsequent applications followed in the work of Mavrakos and Koumoutsakos (1987) and Mavrakos (1991). The major attraction in this scheme is that the boundary value problem for diffraction and radiation is solved together without making any approximations on device size, spacings or the incident wave properties.

According to Twersky (1952), the basic idea while solving wave potentials using MS is as follows. Suppose a plane wave with potential ($\varphi_I$) incidents on a system of freely oscillating members initially kept in rest in the medium. Due to this, a single isolated member (denoted by '$i$') would start oscillating in its all degrees of freedom and hence scatter waves due to its own radiation. This is referred to as first order scattering and can be solved by

$$\varphi_I + {}^1\varphi_S = \psi^1 \qquad \text{Eq16}$$

for the wave radiation problem around body $i$ subjected to the prescribed boundary conditions. Summing over all the devices in the array we can get the total wave potential due to first order scattering as

$$\varphi_I + \sum_i {}^1\varphi_S = \psi^1. \qquad \text{Eq17}$$

In response to this first order scattering but from the other devices (say $i'$) in the array, device $i$ emits second order scattering as

$$\sum_{i'} {}^1\varphi_S + {}^2\varphi_S = \psi^2, i' \neq i. \qquad \text{Eq18}$$

which is a modification of Eqn. (7) by substituting incident potential ($\varphi_I$) with combination of all first order scattering potential (${}^1\varphi_S$) to get the second order scattering (${}^2\varphi_S$). We can proceed in this way upto any arbitrary order (say $m$) and then extending $m$ to infinity,

$$\varphi_S = \sum_{m=1}^{\infty} {}^m\varphi_S \qquad \text{Eq19}$$

we get the total scattered wave field. Then summing up this given order of scattering for all the devices in the array, we can get the total wave field ($\psi = \psi^1 + \psi^2 + \psi^3 + \ldots$). From total wave field other hydrodynamic quantities can be calculated accordingly.

The key task in this scheme is the calculation of scattering coefficients for all orders by satisfying boundary conditions around a given device. In order to accomplish this, the first step is to write the scattering potential () in a cylindrical reference frame ($r,\theta,z$) with origin at the centre of the $i^{th}$ device. Then the boundary value problem is split into various regions: I) $0<r<a_d$, $-d_t<z<0$, $0<\theta<2\pi$; II) $0<r<a_d$, $-d<z<-d+d_t$, $0<\theta<2\pi$ and III) $a_d \leq r<\infty$, $-d<z<0$, $0<\theta<2\pi$ representing cylindrical volume which contains only a single device of characteristic dimension ($a_d$) and draught ($d_t$); beneath the device and the remaining domain respectively. Separate series solutions consisting of unknown coefficients are then written in those regions based on requirement of the boundary conditions (say, Eq4 for wave diffraction problem in region I or Eq6 for wave radiation in all degrees of freedom in zone III). In the next step, the unknown coefficients in the series solutions are obtained by matching them across these regions assuming continuous wave potentials. This technique is more popularly known as matched eigenfunction expansion is what was used by Mavrakos and Koumoutsakos (1987) and their subsequent works (Mavrakos (1991); Mavrakos and McIver (1997)) to solve wave potentials using MS. The resulting expressions involve Hankel and Bessel's functions of first and second kind but various orders given in full details in those papers and hence will not be reported here.

Even though the MS gives exact solutions for the complete wave diffraction radiation problem in the array, its application in practice may be hindered by various issues. Firstly, it is computationally expensive. In theory, any arbitrary order of scattering can be accounted for any number of devices. However, evidences (e.g., Linton and McIver 2001) say the calculations become impractical in terms of computational time for number of devices more than 3 or 5. Secondly, the MS technique assumes that a device is excited by all scattered waves due to other devices uniformly at a given order of scattering (i.e., while calculating second order scattering it is assumed that a device is excited by all first order scattering from other devices). This assumption which holds Eq17 is questionable even with considering the fact that the theory seeks a steady state. Depending on the relative positions of the devices in the array, it is likely that while a device is excited by a higher order scattering from its immediate neighbour, is excited by scattering of lower order from a distant device (or vice versa). Even in the steady state, all the scattering events cannot be uniform as written in Eq17. There should be some provision through which a finite time interval is accounted for the exchange of scattering orders. No experiments have still been found addressing this issue. It is also important to optimize frequency dependent PTO for a single device as well as for the array using MS.

Both of these issues are perhaps better dealt with in the Plane Wave theory described in the following.

### 4.1.2 Plane Wave (PW) theory

The wave scattered from a device can be thought of consisting of two parts: i) a local wave due to evanescent modes and ii) a wave which radiates from the device and acts on another device as a plane wave. Simon (1982) assumed that while the device spacing is larger compared to dimension, the problem can be simplified by ignoring the evanescent part, as it is the later which matters in interaction. In summary, the computation of average $q$ factor ($\bar{q}$) for a linear array of $N$ devices as follows. The incident wave potential with its original phase is expressed within the devices using phase factors (say, $I_1$, $I_2$,...). The effect of coupling is considered by two additional set of terms: i) cylindrical wave potentials ($d_1$,$d_2$,..) originated from other device but measured at a given device using plane wave approximation for the devices lying after the device in hand and ii) similar set of potentials ($c_1$,$c_2$,..) for the devices before the device in hand. By writing these potentials for individual devices in the array a matrix equation is obtained for the unknowns $c_n$ ($n=1,2,..N$) and $d_n$ ($n=1,2,..N$). Then the average $q$ factor ($\bar{q}$) is given by

$$\bar{q} = \frac{1}{N}\sum_{n=1}^{N}|I_n + c_n + d_n|^2 \qquad Eq20$$

The PA theory has been applied so far to investigate interactions among devices in a linear array only. No work has yet been found for other configurations of the array. Indeed, for two dimensional configurations, writing the equations for wave potential for individual devices and then constructing a matrix while accounting mutual scattering is a challenging task. For this reason, the problem can be simplified by looking only at the radiated waves emerging out of the array at the far away boundary ($r\to\infty$). Then it is possible to obtain asymptotic expressions for the radiation coupling terms (i.e., the elements in the radiation impedance matrix) and coefficients for wave excitation forces between devices using Green's theorem and method of stationary phase. These were implemented in the works of Budal (1977) and Falnes (1979) and remains perhaps mostly used theory for calculating interaction factor ($q$) for a WEC array. This is described in the following.

### 4.1.3 PA theory

The term Point Absorption (PA) refers to the fact that the characteristic size of a single device is much smaller than the spacing between devices in the array. Thus, the effect of diffraction can be safely neglected. Using this assumption, Budal (1977) derived the following expression for the $q$ factor:

$$q = \left[1 + \frac{2}{N}\sum_{j=1}^{N}(N-j)\cos(jkL\sin\beta)J_0(jkL)\right]^{-1}, \qquad Eq21$$

where $k$ is the wave number given by $2\pi/\lambda$ and $L$ is the spacing between two devices and $J_0$ is the zeroth order Bessel function. Using this relation, $q\to\pi$ as $N\to\infty$, and $q\to 1$ as $d\to\infty$.

The equal amplitude assumptions for the device oscillations may not be applicable for an array with greater than two devices. Evans (1980) corrected this assumption using

$$q = \frac{1}{N} \mathbf{L} \mathbf{J}^{-1} \mathbf{L}. \qquad Eq22$$

Here, $\mathbf{L}$ is a vector of the form, $\mathbf{L} \equiv \{L_m\}$, $L_m = e^{ikl_m \cos(\beta - \alpha_m)}$, $m=1,2,\ldots n$; $l_m$ is the distance of the device from the origin making an angle $\alpha_m$ with the positive $x$ axis as shown in the plan view of Fig. 1 and $\mathbf{J} \equiv \{J_{ij}\}$ is a $n \times n$ matrix, $J_{ij} = J_0(kL_{ij})$.

In the PA theory, the $q$ factor is independent of the geometry of a single device, i.e. the computed $q$ factor would be same irrespective of whether the devices in the array are cylindrical or spherical. However, one needs a precise description of the geometry of a device in order to calculate its displacement due to incident waves. These displacements are required to check whether the analysis pursued is within the linear wave theory. So, for a given geometry of the device there is a limit in terms of displacements, in order to keep linear wave theory valid. This line of argument essentially reveals that there is a scope to investigate the dependence of the computed $q$ factor over device displacement (i.e. geometry) and seek the optimum displacement for optimum power under the constraints provided by linear wave theory. This was the major motivation behind the work of Thomas and Evans (1981) who computed the conditions for optimum power absorption and optimum displacements for heaving spheres using the PA theory.

Fitzgerald and Thomas (2007) determined that the $q$ factors when integrated over the entire wave incidence angles give unity. Furthermore, Wolgamot et al (2012) has studied this aspect in detail and verified this condition for different array layouts using numerical simulations.

In both in MS and PW there are few more minor assumptions have to be made due to application of the Graff's addition theorem for Bessel function. This results in two minor limitations to the applicability of these techniques to investigating array layouts: i) The vertical projection of interacting bodies on a horizontal plane must not overlap; and ii) A cylindrical coordinate system with an imaginary origin located at the centre of one device, which is required to compute some of the series summations, must not contain the origin of any other device.

### 4.1.4 Comparative studies on PA, PW and MS theories

Some works focused on comparing the performances of the above theories subjected to similar wave conditions, device geometry and array layout. For example,

*a)* Effect of mutual interaction and nonlinear wave forces (Kagemoto and Yue (1986))

They considered the cases of array of cylinders in order to investigate the effects of device spacing, and of evanescent modes on the wave excitation forces on the cylinders. For two fixed cylinders kept at $L/D=2$ and in beam sea ($\beta=90^o$) conditions, the direct matrix method of Kagemoto and Yue (1986) was found to match with the prediction from the Hybrid Element Method (HEM) of Yue et al (1978) for sway excitation force $F_2$. The predictions from zeroth order theory based on phasing only (O) (i.e. keeping the phase of the incident wave intact at the device and using the formula, $F_1(\beta)_0^{th}{}_{-\text{order}}=\cos\{(1/2)k_0L\cos\beta\}F_1$), the theory using no interaction and the PW theory of Simon (1982) without the correction term of McIver and Evans (1984) gives no difference. However, the steady sway drift force ($\overline{F}_2$) causes rather significant mutual-interaction effect.

*Comparison between PA, MS and PW theories (Mavrakos and McIver (1997))*  This important work considered the case of an array of five floating cylinders each of radius $a$ and draught the same as the radius ($a$). Comparative studies were performed for two cases: one with spacing $5a$ and another with $8a$. For these cases, the surge excitation force on device 2 in the array (as shown in the inset of Figure 8a) are shown for spacing $5a$ (Figure 8a) and $8a$ (Figure 8b) respectively. From these studies it was found that PW and MS match better for higher spacing (Figure 8b compared to Figure 8a), because of the assumption in PW that the spacing is more than the wave length. The excitation forces were under-predicted by the PW in the direction of the array for smaller wavelength (higher *ka* in Figure 8a). This may be due to the omission of the evanescent modes in PW. Such modes can be generated as the wave length tends to the device size ($ka \rightarrow 1$). The $q$ factor calculated by the PA, MS and PW theories for both of these spacings are reproduced from Mavrakos and McIver (1997) in Figure 9. It was noticed that for $0.5<ka<1$, the MS and PA matches better than PW. In particular, at the low frequency range, the $q$ factor calculated from PW evidently breaks down due to singularities in the damping matrices (Mavrakos and McIver (1997)). The PW agrees well with MS for $ka> 2$. This was found to be in contrast with the previous case, where even at $ka=0.4$, PW and MS agree well for the excitation forces (i.e. Figure 8a, b). For $ka> 0.8$, PA was found to start differing from MS. This is because in such cases, significant scattering takes place and PA is based on weak scattering assumption. In this range, PW and MS continue to converge for subsequently higher values of $ka$.

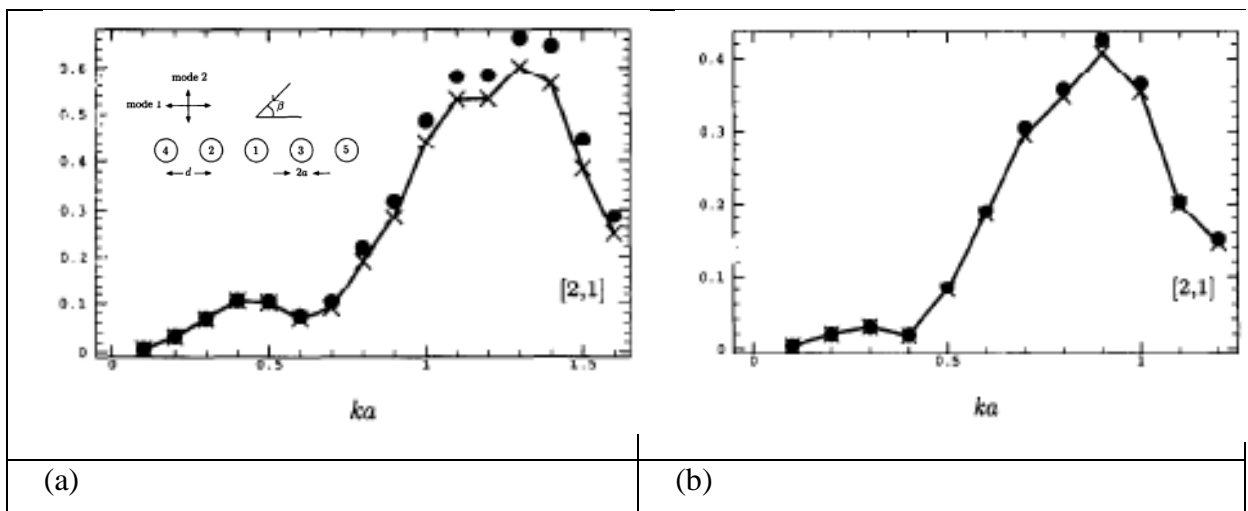

| (a) | (b) |

*Figure 8(From Mavrakos and McIver (1997)) Comparison of the surge excitation force imparted on device 2 in the array (as shown in the inset of (a)) for beam sea computed by PW (•) and MS (×) for (a) spacing 5a and (b) spacing 8a.*

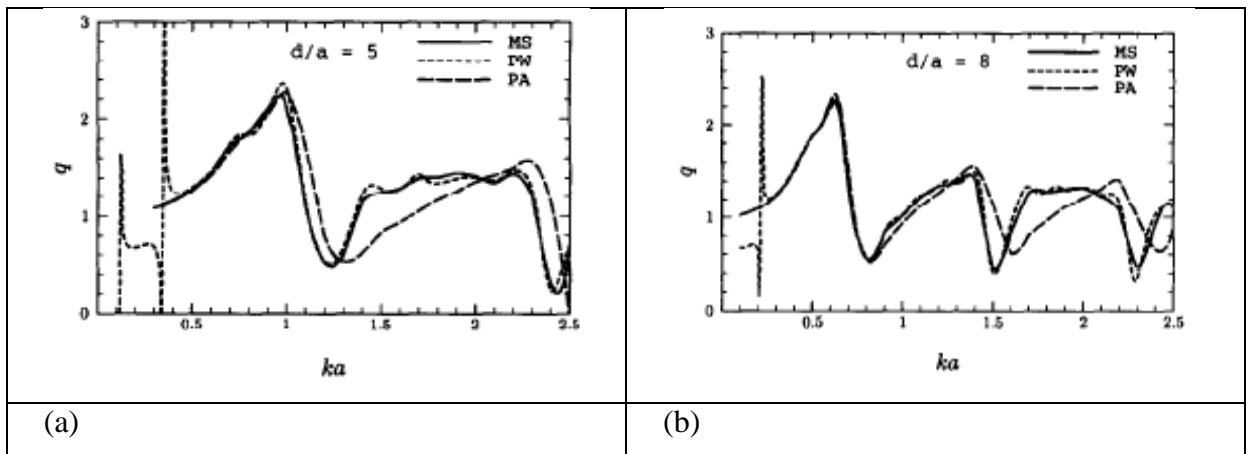

*Figure 9 (From Mavrakos and McIver (1997)).qfactor calculated for (a) spacing 5a and (b) spacing 8a by PW, PA and MS.*

All the above theories are based on deep water approximations ($kd>>1$).Several relations from deep-water approximations, for examplerelations between damping coefficients and excitation force amplitude that represent the far-field for a given device geometry of the device are utilized in the hydrodynamic analysis. Now, if the dimension of the device aremuch smaller than the incident wave length ($ka<<1$), one can apply PA theory, which ignores the effect of diffraction due to the device. If the device is not much smaller than the wavelength, one has to consider diffraction using MS, PW or the Direct Matrix of Kagemotoand Yue (1986). Further, depending on the spacing of the devices compared to the device dimension (i.e. whether $a/L_{ij}<<1$ or otherwise) one can simplify the diffraction analysis by using PW theory ($a/L_{ij}<<1$) instead of employing the MS ($a/L_{ij}\geq 1$) in detail.

## 4.2 Numerical Models of WEC Arrays

A variety of numerical models have been employed to investigate the hydrodynamics of WEC arrays. A survey of these numerical models can be found in Folley et al (2012). They described the analytical models (as described in section 4.1 above) as semi-analytical techniques. This categorization is justified since in practice the infinite series for different expressions are truncated to only a few terms. Hence, the solutions are not exact. However, they can be distinguished from numerical models by considering whether the governing equations are discretised or the approximations are fed directly into them. Based on this argument, this sub section specifically summarizes the numerical works.

These models compute the frequency-dependent added mass and damping coefficient for devices in the array as well as for a single isolated device, under the general assumption of linear wave theory. Here, all the effects of multiple scattering are inherent in the system of equations (Eqns. 1 to 5) discretised using the numerical method in hand. Employing proper resolutions (i.e., grid/ mesh size) all effects of scattering as well as evanescent modes can be captured. Irregular waves or a real sea state can also be reproduced. There are a number of numerical models available for doing this task (i.e., WAMIT, AQUAPLUS, AQUADYN and so on) which use the Boundary Element Method (BEM). It is based on the assumption of potential flow, as are the other analyses reported in the present paper. In this method, the problem is transformed from the entire domain to the boundary using Green's source identity. One sample result is reproduced in Figure 10a) and b) from Borgarino et al (2012), which shows the significant wave height distributions inside the array with the effect of interactions. Nader et al (2012, 2014) carried out similar work but using the Finite Element Method (FEM). Even though such modelling approaches have been employed extensively over past few years for modelling WEC arrays, two serious pitfalls have been noted by the researchers: a) The overall computational cost increases rapidly as the number of devices in the array increases; and b) the water depth is assumed to be constant over the entire computational domain. For this reason, most of studies so far have been limited to a small number of devices. Research to address issue 'a)' has been found in the works of Singh and Babarit (2014) who coupled BEM with the PW theory for fast calculation of scattering by iterations and Borgarino et al (2011, 2012) who used Fast Multipole Algorithm (FMA). The FMA splits the computational surface in a hierarchical oct-tree of panels. The combined effects due to the devices within a panel are then expressed at the centre of the panel using a Green's function multipole expansion. In the next step, information is transferred from the current panel to the target panel using either long- distance or short- distance transfers. The effect of adding one extra device in the arraythenonly requires computing the Green's function multiple expansion around this device. In this way, the complexity of the system is reduced to $O(N)$ compared to $O(N^2)$ in the direct method in which all the devices in the array would have to be considered to compute the far field interactions due to a single device.The assumption of constant water depth may be a good approximation for deep water, but in reality many arrays are likely to be installed in places where there are significant variations in the bathymetry.

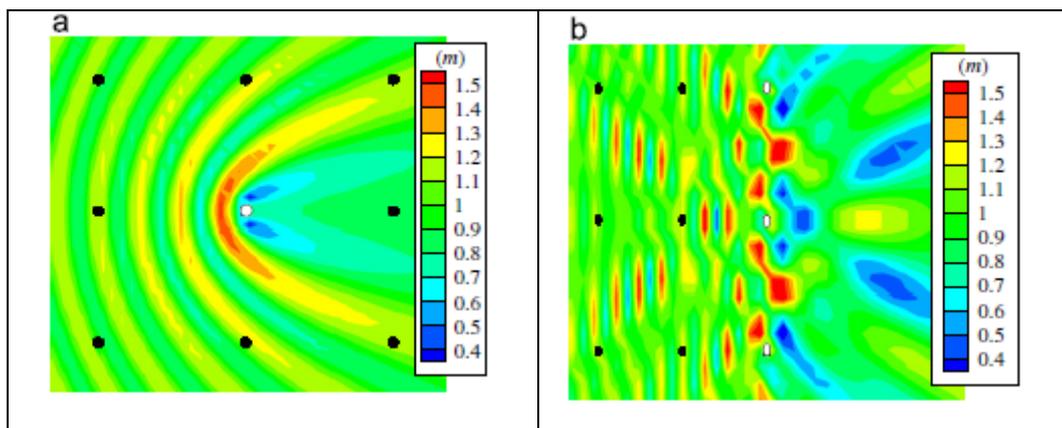

*Figure 10(From Borgarino et al (2012)). Significant wave height distributions for an array of nine heaving devices for incident wave with T= 7.3s (a) around the active cylinder (white) and (b) around three active cylinders in interaction.*

## *4.3 Other theories and methodsapplied to WEC arrays*

### 4.3.1 Phase resolving models:

There are two distinct modelling approaches in this category: i) Boussinesq models and ii) Mild- slope models.

#### 4.3.1.1 Boussinesq models:

Boussinesqmodels are based on a set of nonlinear partial differential equations known as Boussinesq equations. They can describe the wave propagation in the shallow water regime including frequency dispersion and wave breaking effects. The MIKE21 is one such numerical model;Venugopal and Smith (2007) used this model to investigate the performance of an array of bottom mounted devices. Boussinesq models cannot model a floating device. Yet they can be effectively utilized to obtain wave reflection coefficients from the devices in the array. They are also capable of reproducing wave diffraction characteristics effectively.

### 4.3.1.2 Mild slope models:

Mild slope models are based on linear shallow water equations where a water column is represented by a point using depth averaging. They describe the propagation of linear waves over a mildly varying depth with both low computational cost and accuracy. Beels et al (2010a, b) adopted a time-dependent mild slope model named MILDwave to study wake effects past a single and multiple WECs and to study energy absorption. Mild slope based models may be faster and more stable numerically than the Boussinesq models, but in some cases they parametrize the diffraction effect which may actually deviate more than the latter in representing reality. Mendoza et al (2014) used this model to study multiple-purpose devices with two phases: In the first phase, each device was represented as an infinitely high porous box with a finite cross-shore length in the 1D elliptic mild-slope equation model; in the second phase, the 2D modified elliptic mild-slope equation model WAPOQP was used. Both phases are time independent. The mild-slope model was selected because of its simplicity but good accuracy. This project had two constrains: is not applicable under irregular wave conditions; and nor for devices with multiple degrees of freedom.

### 4.3.2 Phase averaging models:

Spectral models come under this category. They work on what is known as the action balance principle. The foundation of such a principle is the conservation of energy. Hence, they seem to be suitable for investigating wave energy absorption in arrays. A variety of physics including wave shoaling, refraction, dissipation due to wave breaking and bottom friction, wind forcing and so on can be studied using such models. They are a natural choice for modelling WEC arrays covering large length scales. However, being phase averaging models, they cannot simulate wave diffraction. This has also been noted by Babarit (2013) while discussing the 'park effect'. The park effect is the effect in which a device in front of another device feels its presence. This feature of an array of WECs is a striking difference from a farm of wind turbines, where only devices in a wake (i.e. behind another wind turbine) experience an effect. Currently, there are two open source numerical models available in this group: i) SWAN, developed by Delft University of Technology; and ii) TOMAWAC developed by Electricité de France. Depending on the grid resolution used, there are two types of spectral models for WEC arrays (Folley, 2012): i) Supra grid models and ii) Sub-grid models.

### 4.3.2.1 Supra-grid models:

Several computational grid points are set over the array. Millar et al (2007) utilized the built-in feature in SWAN for submerged obstacles to represent an entire array. However, it could not give the frequency dependency of the energy absorption. Later, Smith et al (2012) corrected this method by introducing an additional frequency-dependent transmission coefficient. WEC arrays with multiple purposes have been studied recently. Abanades et al, (2014a, b) and Abanades et al, (2015a) combined SWAN and the software XBeach to understand the use of a WEC array as another method to protect the coastline, which is also extracting useful energy from the waves. Abanades et al (2015b) used SWAN and an empirical beach classification (called the Australian beach model), to model the WEC array behaviour and the morphological response of the beach.

### 4.3.2.2 Sub grid models:

Each device in the array is set on a computational grid. The TOMAWAC model allows for both frequency-dependent energy absorption and radiation. Folley and Whittaker (2011) used such a model for a WEC array. Near-field effects cannot be captured properly in this model.

## 4.3.3 Non-linear models:

There are two distinct approaches in this category: i) Nonlinear potential-flow based models and ii) Full Navier-Stokes flow models.

### 4.3.3.1 Potential-flow based models

Similar to the linear wave theory, the Laplace equation is considered the governing equation along with a no-flow condition at the sea bed. However, higher-order nonlinear terms are retained in the combined kinematic and dynamic free surface boundary conditions. This allows modelling of steeper wave amplitudes. Now, similarly to the frequency domain approaches, one can use BEM or FEM to discretise the domain to solve for the velocity potential. Progress in this research can be found in the works of Kashiwagi (2000) for BEM and Yan and Ma (2007) for FEM. There are few evidences in literature (e.g., Yan et al. 2012; Ma et al. 2013) which shows the application of non- linear potential flow based model in time domain simulation of multiple floating bodies in waves. However, till now no such papers are found applied to wec array.

### 4.3.3.2 Full Navier-Stokes flow models

These comprise the codes commonly described as 'Computational Fluid Dynamics' (CFD). These models solve continuity and momentum conservation equations for the fluid domain. In contrast to the potential flow based models, viscous and nonlinear effects are inherent. Turbulence can be dealt with either by Direct Numerical Simulation (DNS), or Reynolds Averaged Navier Stokes equation (RANS). In DNS, all length scales down to the viscous scale are computed, which can be computationally prohibitive for realistic conditions. In RANS, the Navier Stokes equations are solved over given length scales that are computationally tractable, but a sub-gridscale model of the turbulence is required. One can adopt numerical methods like the Finite Volume Method (FVM), Mesh-free methods like Smoothed Particle Hydrodynamics (SPH) etc. for discretisation purposes. Both FVM and SPH discretization methods carry the capability of multiphase simulations, allowing for both air and water motions to be resolved. Examples of such models are the commercial code AnsysFLUENT, CD-AdapcoStarCCM+ and open-source solvers like OpenFOAM. Time domain models have been successfully applied and developed for calculating nonlinear wave loads on fixed, floating or elastic structures. However they become prohibitive in terms of computational resources while modelling both near and distant effects (i.edue to meshing requirements). This barrier can be overcome with three strategies. Firstly, an efficient computational strategy in a parallel environment (for example, MPI, OpenMP or GPU computing) can be adopted. Secondly, variable resolution can be used, in which there is higher resolution near the point of interest compared to the rest of the domain.; Thirdly, hybrid coupling between two different solvers suitable for different requirements can be used, for example, Josset and Clément (2007) and Sriram et al (2014). Recent applications of CFD codes for modelling WEC devices include Luo et al (2014), Henry et al (2013), Wei et al (2014, 2015) and Zhang et al (2012). There are few applications of CFD codes for full scale WEC arrays; Agamloh et al. (2008) on an array of two devices and the work of Westphalen et al (2009) are amongst a few such examples.

Considering all these various modelling approaches (analytical, numerical), it becomes apparent that there is no single method which is best suited for all investigations of WEC arrays. Pushing computations to resolve more detail within the flows is demands increasing computational resources. Each modelling approach possesses its own inherent limitations which can be better dealt with by another approach. The complementary nature (for example, the assumption of potential flow) underlying the different approaches are probably best utilized by developing suitable numerical schemes where different solvers can run simultaneously through common interfaces shared by all.

# 5 Applications

There have been a number of important works carried out for solving velocity potential depending on several other functional requirements for the devices in the array (As mentioned while discussing the linear wave theory (Eqns. (1) to (4)). This essentially reflects the efforts undertook by various companies (as discussed in Part 3) in extending their existing different prototype for single device towards array. Few of them are discussed in the following.

## *5.1 Effect of Power Take Off (PTO) control*

Falcão (2002) presented a theory dealing with an infinite linear array of heaving devices with Power Take Off (PTO) control by a turbine. The effect of such control was reflected in the array efficiency $\hat{\eta}$ (which in fact is related with the capture width ($l_{max}$ in Eqn. 6) by $l_{max} = \hat{\eta} b \cos\beta$) as shown in Figure 11a and b. Actually, Falcão and his co-authors reported many works (i.e., Henriques et al., 2012) which investigated the dynamics of PTO control of a WEC array. Works on similar topics can also be found in Price et al (2009) and Babarit et al (2009).

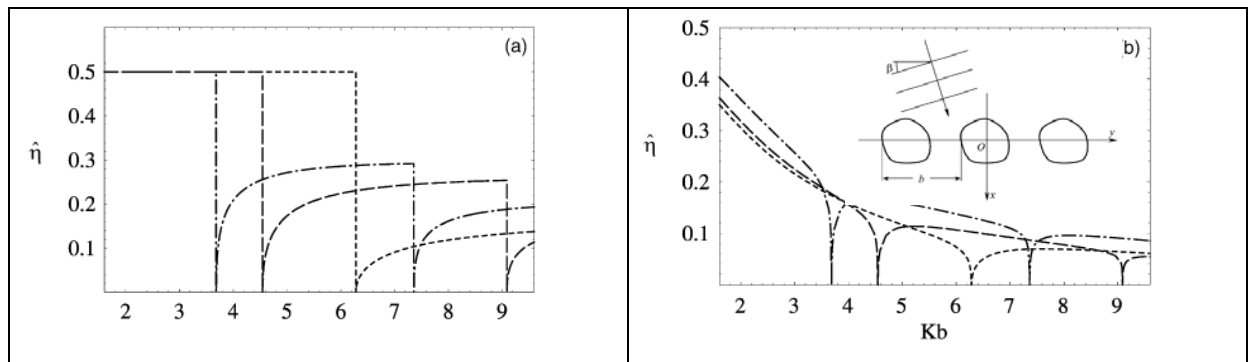

*Figure 11 (From Falcao (2002)). Effect of turbine PTO on efficiency of array (as shown inset of (b)) for β=0 (..), π/4 (--) and π/8 (-.-), (a) with control; (b) without control.*

## *5.2 Array of line absorbers*

Stansell and Pizer (2013) developed a theory using the relative motion hypothesis for investigating capture width of attenuator type wec devices under the constraint of constraint of their own swept volumes. An attenuator type device has both surge and heave motions. In this case, a leading order relationship between diffracted and radiated wave potential can be

obtained.Higher capture width were shown to be possible for attenuator type wecdevcies compared to point devices.

## *5.3 Positioning*

### 5.3.1 Effect of Distance

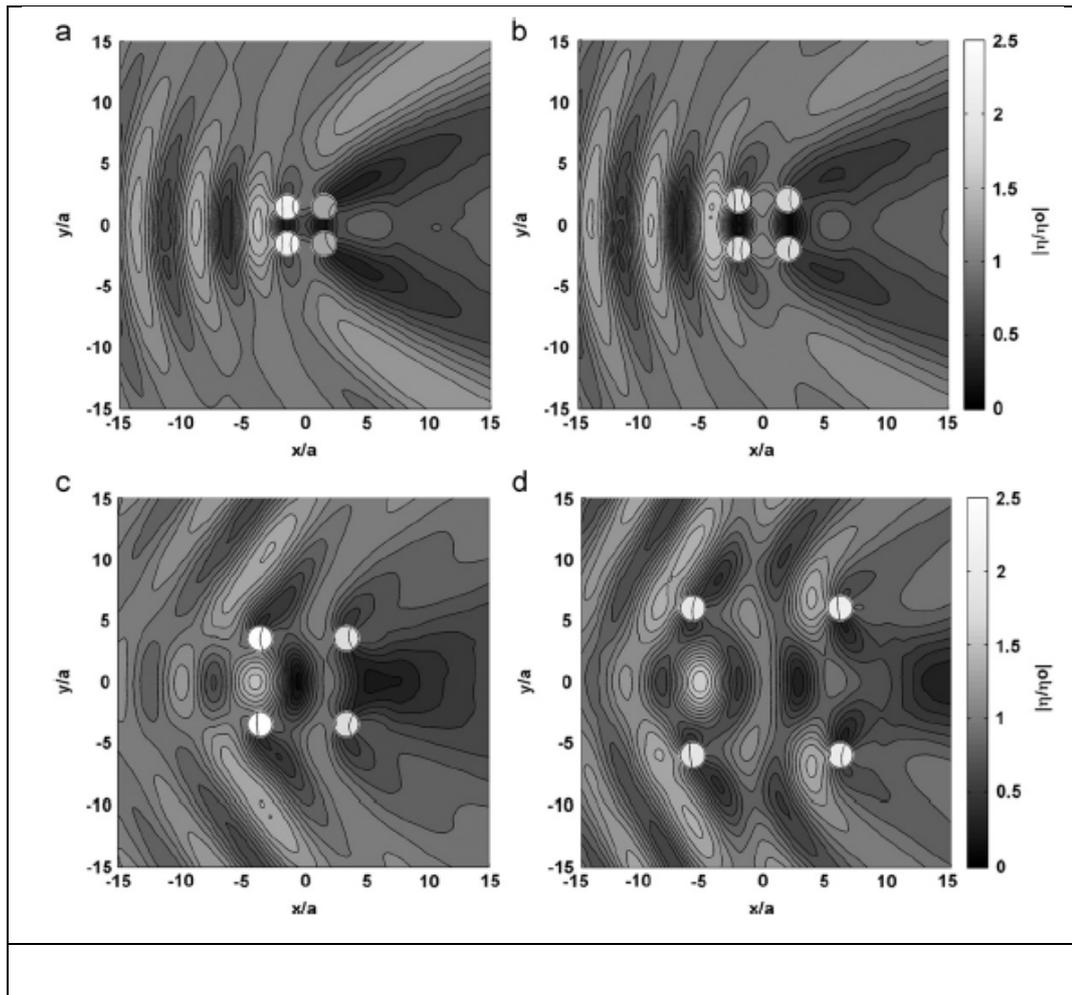

*Figure 12from Nader et al (2012)). Total wave amplitude around and in the four OWC device array for a spacing of (a) L/a= 1, (b) L/a= 2, (c) L/a= 5 and (d) L/a= 10. The angle of the incident wave is 0 and the frequency is* kh= *3.4. Here,* L *is the array spacing.*

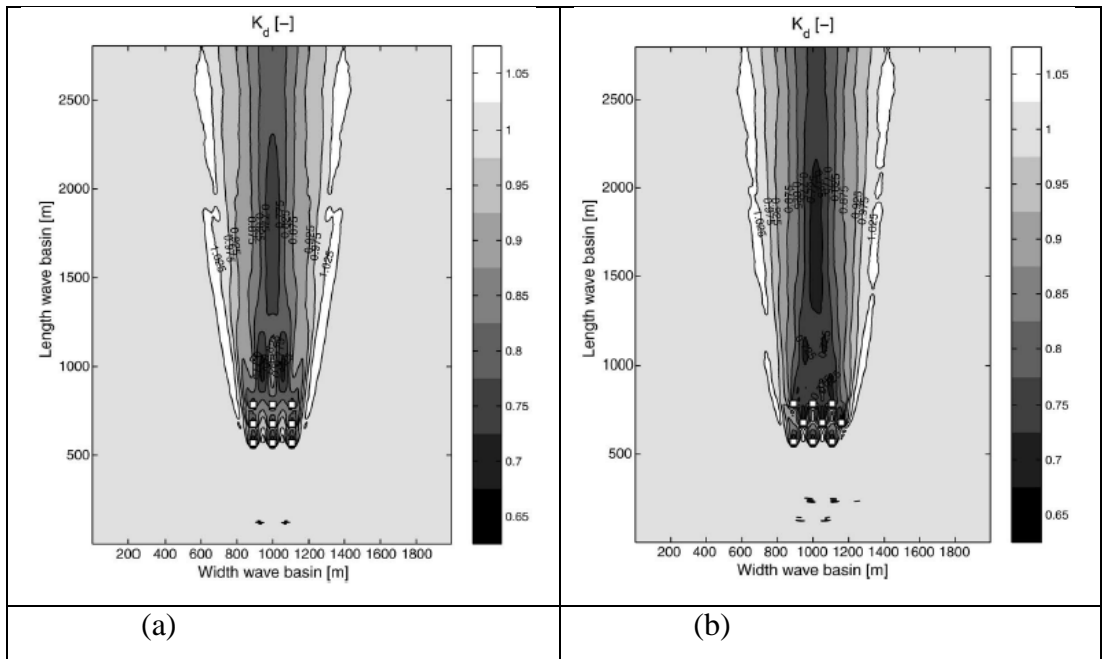

*Figure 13Disturbance coefficient for an array of nine devices for (a) regular and (b) staggered orientation under irregular wave of $T_p$= 5.2s.*

### 5.3.2 Bragg scattering and entrapped modes within the array

The effect of entrapped modes around periodic arrays of cylinders has been extensively studied using linearized theories and had identified many symmetric and anti-symmetric trapped modes for vertical cylinders (Callan et al,1991; Porter and Evans,1999;Maniar& Newman,1997). The excitation of these modes near the array resonance frequencies may pose a danger to the structures and is therefore an engineering concern. The effect of scattering waves away from resonance has been treated by Linton & Evans (1990,1993) andMcIver & Bennett (1993) and resembles the theory of diffraction gratings. Nonlinear theories for the these resonating modes for both monochromatic and narrow band spectra incident waves have been given by Sammarcoet al. (1997a,b), who showedboth theoretically and experimentally that in the former case resonance becomes chaotic. Working with non-linear theories, it was shown that these trapped modes can be excited subharmonically by monochromatic incident waves of twice the Bragg resonance frequency.Sammarcoet al. (1997a,b) investigated the growth of the resonance amplitudes when forced by narrow-band incident wave spectra. Kagemotoet al. (2002)experimentally investigated these resonating modes near the critical resonance frequency for long cylindrical arrays and found deviations from potential-flow theory at particular incident wave period, which was attributed to viscous effects the array structures.The effect of Bragg scattering in an array of WECs was also studied by Garnaud and Mei (2009). Studies related to the understanding of the similar entrapped modes inside the array can also be found in Chen et al (2012).

### 5.3.3 Optimisation

The effect of array layout or configuration on the performance of an array of WEC device has been addressed in the work of Folley and Whittaker (2009), Child (2011), Child and Venugopal (2010). Using linear wave theory, they developed a model where the effect of array configuration on power extraction can be studied using an optimization algorithm like the Parabolic Intersection (PI) and Genetic Algorithm (GA) methods. Both regular and irregular wave conditions were considered. One sample result showing the optimization of array configurations provided by GA is shown in Figure 14.

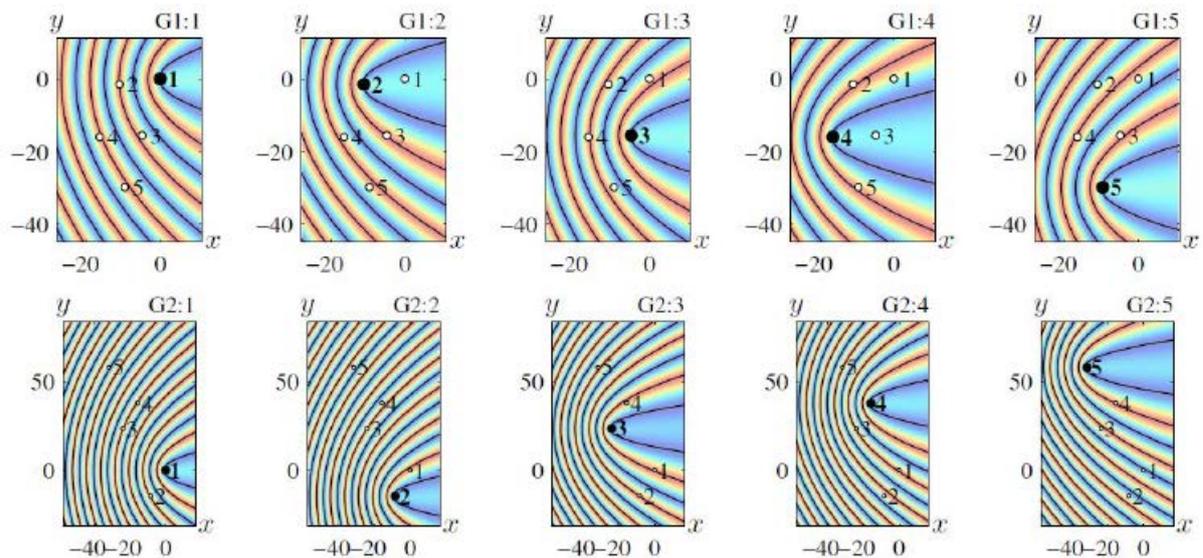

*Figure 14 (From Child and Venugopal (2010)). Relative phase between the originated and the outgoing waves. In the colourbar, red corresponds to in phase and blue corresponds to out of phase with respect to the originated wave. The array layout is chosen by GA. The point filled with black is the source of the wave.*

The $q$ factors obtained by GA and PA for similar conditions are provided in Table 2, reproduced from Child and Venugopal (2010). Details of the array layouts are shown in Figure 15. Here, G1, G2 and G3 represents array layout obtained from GA and P1, P2 and P3 represents array layout obtained from PA methods.

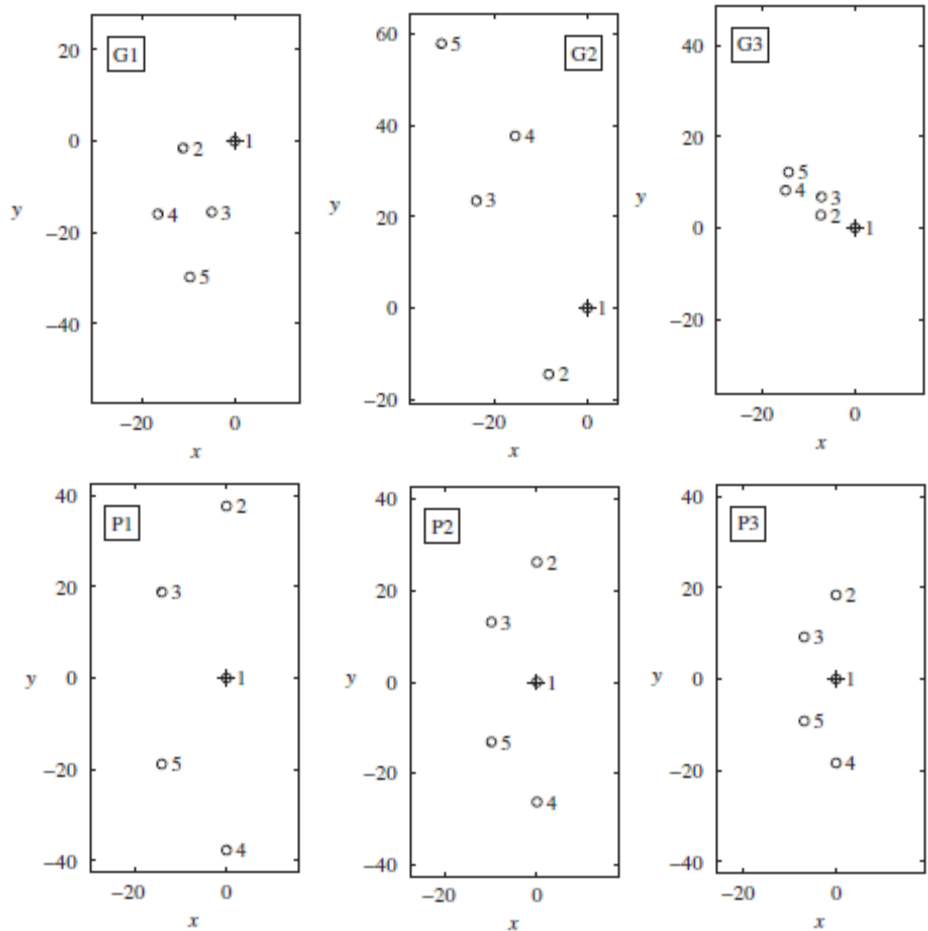

*Figure 15 (From Child and Venugopal (2010)). Details of the adopted array configurations.*

*Table 2 (From Child and Venugopal (2010)). Computed array interaction factors in different array layouts obtained by GA and PI.*

| Array with GA | q factor | Array with PI | q factor |
| --- | --- | --- | --- |
| G1 | 1.163 | P1 | 1.136 |
| G2 | 2.010 | P2 | 1.787 |
| G3 | 0.326 | P3 | 0.453 |

## 5.4 Effect of mooring

The effect of a mooring system on the performance of an array of three hemispherical WEC devices was studied in Vicente et al. (2009).

Modelling of such a system requires writing the equations of motions in time domain in each degree of freedom of the devices. The work of Vicente et al (2009) is one of very few paperswhich adopt such an approach for WEC arrays. Yet, if one wishes to take into account the nonlinearities inherent in the system either in the wave motion, in the mooring or in the PTO, a time domain analysis is indispensable. This was first adopted by Cummins (1962) for ships and later extended to WEC devices by Jefferys (1980). One key task here is the dealing with the convolution term for radiation impulse response function. However, since this convolution is linear it can be substituted with a linear time invariant system like a transfer functions or a state space system. System identification is required for this purpose. System identification is a technique to account for the convolution integral in terms of known hydrodynamic quantities.

## 5.5 Theory for array of buoys of small fixed structures

Garnaud and Mei (2009,2010) developed a theory for an array of small buoys fixed in a rig structure. The initial design was proposed by Fred Oslen and ABB Power Systems (www02.abb.com) in Norway and also adopted by the Manchester bobber (www.manchesterbobber.com). This work analysed the performance of the array as it runs hydraulic PTOs during heaving due to the incident waves. The coordinate system which defines the device dimension and spacing (micro scale) in the array is related with the coordinate system which defines water depth and wave length (macro scale) using coefficient in linear expressions. The entire boundary value problem is considered in terms of order of the coordinate scales: leading order (within a micro scale unit with homogeneous forms) first and second orders (inhomogeneous). By homogenization of different scale effects, they were able to compute the required wave elevations and phase information both inside as well as distant from the array.

## 5.6 Array of flap type WECs

Renzi and Dias (2012a) developed a theory for a periodic array of flap-type WECs based on their earlier work (Renzi and Dias, 2012b). An example of this class of WEC device is the Oyster technology by Aquamarine Power noted in section 3. Details of the Oyster technology can also be found in Renzi et al (2014). Sarkar et al. (2014) extended the work of Renzi et al (2014) to investigate the performance of a large array of flap type WEC devices. In this work they used a modified *q* factor proposed by Babarit (2010), $q = (P_i - P_{single})/\max(P_{single})$, $i=1,2,..N$ instead of the basic definition $q = P/NP_{single}$ (as mentioned in the beginning of this chapter), in order to better understand the performance of a single device in the array. This shows the situations in which there is no significant gain in the *q* factor from increasing the number of devices in a particular layout.

### 5.6.1 Non-linear effects

# 6 Experimental Models of WEC Arrays

Since Salter's paper was published in 1974, many experimental projects have been working on WEC prototypes, as well as their economic issues, identifying regions of energy resources, numerical models, and their environmental issues.

Experiments have an enormous relevance to developing wave energy projects. The reason for the experimental tests is to validate the information obtained from the multiple numerical models developed until today. WEC arrays are not limited to power generation waves. Some prototypes or arrays have been used as a source of fresh water (Sharmila et al, 2004) and for coastal protection (Zanuttigh and Angelelli, 2013), as previously pointed out in section 3 while discussing the activity of OceanLinx. Furthermore, a project reported in recently by Perez-Collazo et al (2014) aims at combining wind- and wave-energy energy extraction. This section was sub-divided in accordance with experimental test locations and some features of the devices.

## *6.1 WEC arrays extracting energy, laboratory cases.*

The laboratory experiments have many variations, mainly due to varying project objectives. Here the reported works will be discussed based on the mode of operation of WECs in the array: a) point absorber including heaving buoys; b) OWCs and c) overtoppinger.

### 6.1.1 Point Absorber / Heaving buoy type:

A project described in Ashton et al (2009), was tested at Trondheim, Norway, in the MARINTEK's wave tank. During the project a generic oscillating water column model was used at 1:20 scale. The PTO was represented by a small orifice in the top of the buoy. A total of five devices were used and performed in a range of regular and irregular waves, generated from a Bretschneider spectrum. Also, 6 probes were tested, with three different arrays: a single device, a 3-device and 5-device array. A measurement package installed on the devices included a motion tracking system. An unexpected increase in power capture was reported when multiple devices were installed. An increase in wave power at the third probe, which showed the largest power, was explained by Ashton et al (2009), who suggested that hydrodynamic effects dominate the power capture in the measured near-field wave states. This was also previously found by Mavrakos and McIver (1997) while analysing the hydrodynamic forces on a device in a array using PA, MS and PW theories (see section 4.1.4). Spatial variability around this array has been seen to be significant, demonstrating that the influence of the placement of point wave sensors on the wave field measured must be considered when analysing wave measurements for impact studies. The authors also suggest a further study which could include a full hydrodynamic analysis of the body, in order to understand the contribution of complex interactions within the array.

The project undertaken by Haller et al (2011) was developed at the Tsunami Wave Basin (TWB) at the Hinsdale Wave Research Laboratory within Oregon State University. Experiments were accomplished with five WEC devices (point-absorber-type heavingbuoy) at a scale of 1:33, using a commercial design, "Manta" from Columbia Power Technologies. Wave conditions considered during the experiment were regular waves and real sea (irregular) waves. The experimental instrumentation consisted of wave gages which were installed in three separate lines, current meters, and a 3D WEC motion-tracking system. Several different array configurations were tested. The hydrodynamic observations were extensive. Initial results quantified the degree of wave shadowing induced by the various incident wave conditions and demonstrate the dependence of the shadowing on the chosen array configuration. A more closely-spaced array clearly shows more shadowing. This is in good agreement with the findings from the numerical simulation of the mild slope phase resolving model of Beels et al (2010a) (refer to section 4.2 Numerical). Frequency and directional spreading in the wave field smooths the wave height variation and reduces the shadow. Motion tracking results show an association between wave shadowing and device performance, which indicates that wave absorption, not scattering, is the dominant process here. Use of these results could allow WEC array designer to consider environmental aspects.

A further project was developed at the Oregon State University (Porter, 2012), using the same methodology as reported by Haller et al, 2011. However, more recent version of the WEC prototype (Manta 3.1) was used. In addition, bathymetric surveys were made before and after the experiment, in order to measure changes to the beach due to the experiments. The experimental data set is large and has not been fully utilised to date. The lee of the array has a shadow with respect to the incident wave conditions, due to the real seas applied and regular trials show a great influence of frequency and array size. The scaling from a three-device to a five-device array is moderately linear for both real-sea (irregular) and regular waves. After the comparison between the predicted shadow (numerical model) and measured shadow results, a discrepancy was found. This difference was found to be due to scattering short waves caused by additional hydrodynamic interactions of the WEC with the incident wave field (Porter, 2012).

The project conducted by Stratigaki et al, 2012, was carried out in two different laboratories: the wave flume of Ghent University (Belgium), and the wave basin of the Queen's University of Belfast. The main aim of this study was to determine experimentally near-field and far-field wake effects from large WEC farms composed of oscillating point absorbers, under different array configurations and wave conditions. The methodology utilised in both locations included regular, irregular and short crested waves, and four different test cases: 1) without the presence of the buoy; 2) with the buoy as a fixed obstacle; 3) with the buoy but no damping applied through the PTO; and 4) with the buoy and damping applied through the PTO. The PTO was simulated by a simple mechanical brake. The experiment started with one device, and gradually more devices were added. The Ghent model was a simple and easily reproducible structure, as well as a good structure stability test. The Belfast project showed the efficiency and repeatability of the developed PTO system. This project could be considered a preparation step for the further tests in the 3D shallow water wave basin at DHI in Denmark (Troch et al, 2014). Even though only one device type was used, the importance of this project is as the basis of the biggest experimental test ever carried out.

To date, the project reported by Troch et al (2014) is considered the largest experimental test set-up of WECs with a total of 25 individual units (heaving buoys) in an array. The experiment was performed at the DHI Shallow Water Wave Basin (Denmark). Troch et al (2014) reported that no prior experimental studies are publicly available detailing the WEC response, power output and wave field modifications due to an array. The objectives were to find the impacts on power absorption and on wave conditions. Four types of waves were considered: regular, polychromatic, long-crested irregular and short-crested irregular waves. The array tested was rectilinear, composed of $5 \times 5$ (heaving buoys with 12 different configurations. The results clearly show wave attenuation and a significant wave-height decrease due to the power extraction. The wave amplitude attenuation data could be used for estimation of coastline evolution due to the presence of the WEC array. This data base can be used to validate and extend a large range of numerical models used to model the response, power absorption and wave field modifications due to oscillating WECs. The data base could be extrapolated to floating structures and platforms, or to stationary cylinders under wave action, etc. (Troch et al, 2014).

It is important to note a recent large study, the WECwakes R & D Project funded by the EU FP7 HYDRALAB IV[8], which at the time of writing is still in the process of being documented in the archival literature.

### 6.1.2 OWC type:

In 2008 a project was developed at the NTNU Ocean Basin in Trondheim, Norway (NTNU Ocean Basin). As a result of this investigation two papers were published at 2010. Methodologies, reported by Krivtson and Linfoot, (2010) and Bryden and Linfoot, (2010), were conducted using five sets of experiments (based on its position on the array, each device was identified with a number, 1-5). The sets were 1) Basin calibration (no device); 2) Tests on WEC 1 in regular and irregular waves and currents; 3) Tests on five WECs (1-2-3-4-5) in regular and irregular waves and currents; 4) Tests on three WECs (1-2-3) in regular and irregular waves and currents; and, 5) Tests on the damping and mooring stiffness of WEC1. Moreover, three different arrays were studied: a single device; three devices; and five devices.The WECs tested were generic OWC devices each fitted with an adjustable damping orifice plate. Krivtson and Linfoot, (2010) provided a preliminary analysis of the extreme mooring loads using a 1:20 scale single device. The results are broadly consistent with the hypothesis that the mooring loads should differ. The result in short crested sea conditions indicate that peak mooring loads in a multi-WEC array may be considerably higher than in a single-WEC configuration. Similar findings were also reported by Vicente et al (2009), who performed numerical simulations on effect of mooring (refer to 4.1.3). It was found that the power capture of WEC1 was considerably enhanced in a 5-WEC array compared with a single WEC. This was also found by Ashton et al (2009), mentioned above. This effect was explained by Krivtson and Linfoot, as additional capturing of the incident wave energy, with its consequent release by radiative damping. The experiments reported by Krivtson and Linfoot (2010), showed that the testing of mooring systems for WEC arrays should include short crested seas so that the hydrodynamic interaction between the devices can be quantified. The extreme peak mooring loads in the leading mooring line were approximately doubled relative to those in a single device under similar environmental conditions. Bryden and Linfoot (2010) briefly described the physical model. To model the PTO from OWC devices, they were fitted with an adjustable damping orifice plate. As was expected, the analysis of the data indicates that close-packed arrays of energy converters may be more efficient in energy capture than the same number of individual devices.

---

[8] http://www.ugent.be/ea/civil-engineering/en/research/coastal-bridges-roads/news-events/news/wecwakes-project.htm

### 6.1.3 Overtopping type:

In 2011, Magagna et al carried out a project at a scale of 1:40 investigating an array of three onshore OWC wave pumps (OWCPs). This prototype can be considered as an overtopping type of WEC; however it differs from the standard overtopping devices such as the Wave Dragon or the Composite Sea Wall, since OWCPs exploit the run–up of the water over an inclined ramp to deliver water to a reservoir. Each device used a different inclination-ramp angle (25º, 30º and 35º). The separation distance between the devices played an enormous role during the project and was evaluated through four different geometrical configurations of the array. This importance was also noted by Monk et al (2013) as discussed in section section 1.2. The PTO system was tested for each device. In particular, the central device (ramp angle of 30º) was positively affected by its position in the array, with the side device achieving lower efficiency. The values of $q$ were determined both in terms of delivered mass and efficiency, with higher values obtained for the case with higher separation distance. In particular, $q$ was found to be dependent on the wave frequency. Each device had differences in the phase of response, suggesting that whilst the wave forces the movement of the devices simultaneously, their radiation component is differently phased, reducing the destructive component of the radiation. Values of $q > 1$ and values of $q_e > 0.33$ were obtained, showing that a positive effect between each array component can be achieved if the separation distance between the devices is optimized.

## *6.2 Current research*

## *6.3 WECs arrays extracting energy, real conditions.*

To date, there are not too many projects in which prototypes are tested in open ocean locations, apart from the full-scale deployments described in section 3. The implementation of projects in real ocean conditions is limited, owing to the high risk the prototypes are exposed to natural sea conditions, and the high cost of installation. Furthermore the funding required for open-ocean experiments is in most of the cases far greater. These studies have to be carried out carefully to avoid possible partial damage or loss of whole project during the test. However, such experiments offer the genuine real interaction between devices under natural ocean conditions.

Rahm et al (2012) studied the aggregation of power from two and three WECs, and quantified the level of smoothing. During this study the standard deviation of electrical power as a function of various parameters was investigated. The site was Lysekil off the Swedish West Coast. The first WEC was deployed in 2006; additionally two more WECs were installed in 2009. In the longer run in 2009, three WECs were operated over 19.7 consecutive days. The exact damping functions used during the experiment are unknown. It was not possible to detect any influence from wave direction on the standard deviation of electrical power. The standard deviation reduction was found to only be due to the increase in significant wave height..

## 6.4 WECs arrays with multiple applications

The main objective of the WECs is the extraction of power from ocean waves. Nevertheless, as aforementioned, some WECs are multiple-purpose devices (Abanades et al,2014 a, b;Abanades et al, 2015a, b). One of those applications is as an alternative coastal protection structure that meanwhile, is extracting energy from the waves.

In 2013, Zanuttigh and Angelellistudied WECs as an alternative technique to protect the coastline that are also extracting energy from the waves. This research was undertaken at the Aalborg University (Denmark) facilities. The devices were "DEXA" models classified as Wave Activated Body types with a scale of 1:30 for a single device and 1:60 for three identical devices. The DEXA device consists of two rigid pontoons with a hinge in between, which allows each pontoon to pivot in relation to the other. The PTO system was placed close to the centre of the system, in order to maximise the stabilisation force (Zanuttigh and Angelelli, 2013). The wave pattern applied considered of regular, irregular short-crested and irregular long-crested waves. Oblique 2D and 3D waves were generated. The results were described in terms of wave transmission, wave reflection, change of wave direction induced by the device and mutual interaction among the devices. The effects of the device on sediment transport processes, and the changes in wave direction behind the device were evaluated.. These results show that these devices lead to a modest reduction or absorption of the incident wave energy; the transmission coefficient ($K_T$) was always greater than 0.8 for the single device at 1:30 scale and always greater than 0.65 behind a two-line staggered wave farm at 1:60 scale. The devices induce a change of wave direction of a few degrees which should be carefully considered when examination of the effects such installations may produce on near-shore sediment transport processes. The water depth at the installation did not significantly affect the induced hydrodynamics, leading to the conclusion that a farm of these devices would not be particularly sensitive to sea level rise. The authors gave some consideration to improving the WEC experiment, concerning the device design and the array layout. With regard to the device design, it was reported that the device length has to be tuned on the basis of the local peak wave length, and that a heavier system will give a better reduction in the incident wave energy. It was suggested that the benefits of the interaction between the devices included: reducing the device gap, permitting the devices and moorings to freely move; economic optimisation of the array having the devices staggered, so that the devices are still reached by a great amount of residual wave energy; and modifying the configuration to be repeated in the cross-shore direction to provide an appreciable sheltering effect.

# 7 Conclusion

The numerous advantages of having an appropriately configured array of wave energy converters instead a single device have been established by different studies, beginning with purely analytic theory in the 1970s, and progressing to numerical modelling and experimental modelling in recent years. Conversely, disadvantages of an inappropriate configuration have also been revealed. In addition to the studies reported herein, some of theoretical and numerical calculations will have been commenced within some of the wave-power companies. However, commercial realities mean that such calculations would be case-specific, and underlying principles of array operation are not yet perfectly understood. For example, one of the most serious problems in the development and laboratory validation of any theory modelling of Wave-Energy Converters (WECs) is the loss of energy from individual devices due to turbulence. The reciprocating nature of the flow created by waves means that the nature of this turbulence is completely different from that of the uni-directional turbulent flows which have been successfully modelled for over 70 years. Hence, it is essential to undertake laboratory results on model WECs at more than one scale, before assuming the results can be scaled up. This applies to arrays of WECs as well as to individual machines.

The experimental studies on WEC arrays undertaken so far have captured the inherent physics of the array system. The fundamental attributes of array behaviour have been predicted by the existing analytical theories or the more recent numerous numerical models developed. Thus, for the limited range of situations within the capabilities of linear wave theory, the present state of the art modelling technique is reasonably adequate. So far, few reported theoretical or numerical modelling studies utilise to reproduce physical model data.

## Acknowledgement

This work is supported by the Australian Renewable Energy Agency, Emerging Renewables Program grant A00575.